\documentclass[article,twocolumn,showpacs]{revtex4}

\usepackage{graphicx}% Include figure files
\usepackage{epsfig}
\usepackage{xspace}
\usepackage{float}
\DeclareMathSymbol{\Delta}{\mathalpha}{letters}{"01}
\DeclareMathSymbol{\Sigma}{\mathalpha}{letters}{"06}
\DeclareMathSymbol{\Lambda}{\mathalpha}{letters}{"03}
\DeclareMathSymbol{\Xi}{\mathalpha}{letters}{"04}

\newcommand{\rh}{\ensuremath{\rho(770)^{0}}\xspace}
\newcommand{\ks}{\ensuremath{K^{*}(892)^{0}}\xspace}
\newcommand{\aks}{\ensuremath{\overline{K}^{*}(892)^{0}}\xspace}
\newcommand{\ph}{\ensuremath{\phi(1020)}\xspace}
\newcommand{\dl}{\ensuremath{\Delta(1232)^{++}}\xspace}

\newcommand{\ssx}{\ensuremath{\Sigma(1385)^{\pm}}\xspace}

\newcommand{\ls}{\ensuremath{\Lambda(1520)}\xspace}

\newcommand{\xs}{\ensuremath{\Xi(1530)^{0}}\xspace}

\newcommand{\pb}{Pb-Pb\xspace}

\newcommand{\rsnn}[1][2.76~TeV]{\ensuremath{\sqrt{s_{NN}}=}~#1\xspace}

\newcommand{\dnc}{\ensuremath{\langle dN_{\mathrm{ch}}\kern-0.06em /\kern-0.13em d\eta\rangle}\xspace}
\newcommand{\dncr}{\ensuremath{\dnc^{1/3}}\xspace}

\newcommand{\pT}{\ensuremath{p_{\mathrm{T}}}\xspace}
\newcommand{\mpt}{\ensuremath{\langle\pT\rangle}\xspace}
\newcommand{\gvc}{\ensuremath{\mathrm{GeV}/c}\xspace}

\begin{document}

\hyphenpenalty=100000

\title[]{Hadronic resonance production and interaction in partonic and hadronic matter in EPOS3 with and without the hadronic afterburner UrQMD}

\author{A. G. Knospe$^1$, C. Markert$^1$, K. Werner$^2$, J. Steinheimer$^4$, M. Bleicher$^{3,4}$}

\affiliation{$^1$ The University of Texas at Austin, Physics Department, Austin, Texas, USA \\
%Department of Physics, University of Texas at Austin, Austin, TX 78712, USA \\
$^2$ SUBATECH, UMR 6457, Universit$\acute{e}$ de Nantes, Ecole des Mines de Nantes, IN2P3/CNRS. 4 rue Alfred Kastler, 44307 Nantes CEDEX 3, France \\
$^3$ Institut f\"ur Theoretische Physik, Johann Wolfgang Goethe-Universit\"at,
Max-von-Laue-Strasse 1, 60438 Frankfurt am Main, Germany\\
$^4$Frankfurt Institute for Advanced Studies, Ruth-Moufang-Strasse 1, 60438 Frankfurt am Main, Germany}

\vspace*{1cm}

\begin{abstract}
We study the production of hadronic resonances and their interaction in the partonic and hadronic medium using the EPOS3 model, which employs  the UrQMD model for the description of the hadronic phase.  We investigate the centrality dependence of the yields and momentum distributions for various resonances (\rh, \ks, \ph, \dl, \ssx, \ls, \xs and their antiparticles) in \pb collisions at \rsnn. The predictions for \ks and \ph will be compared with the experimental data from the ALICE collaboration. The observed signal suppression of the \ks with increasing centrality will be discussed with respect to the resonance interaction in the hadronic medium. The mean transverse momentum and other particle ratios such as \ph/p and ($\Omega+\bar{\Omega}$)/\ph will be discussed with respect to additional contributions from the hadronic medium interactions.

\end{abstract}


\maketitle

\section{Introduction}

At RHIC and LHC energies a state of partonic matter consisting of quarks and gluons, the Quark Gluon Plasma (QGP), is expected to be created in high energy heavy-ion collisions. As the system expands and cools %ch take out "below a" and replace it with "down to the"
down to the critical temperature $T_{\rm c} = 155-160$~MeV \cite{lattice_wu,lattice_bnl} a transition from partonic matter into hadronic matter will occur. After hadronisation, the system continues to expand and interact hadronically.  The standard description assumes two freeze-out conditions in the hadronic phase: the end of inelastic interactions at the chemical freeze-out temperature $T_{\rm ch}$ at which the abundances of stable particles are fixed and the end of elastic interactions at the kinetic freeze-out temperature $T_{\rm kin}$ at which the final particle \pT spectra are determined (frozen).

Due to their short lifetimes of a few fm/$c$, the interactions of hadronic resonances span the full expansion of the collision including the partonic and hadronic phases \cite{Markert_Vitev}. Resonances will dissociate and regenerate in the partonic medium and decay and regenerate in the hadronic medium. 
%ch  new text maybe
%If there is a parton hadron mixed phase around the phase transition or if resonances are formed earlier due to their heavy masses \cite{Markert_Vitev}, resonances will also decay and regenerate during the late partonic and mixed phase.

Therefore resonances are a unique tool to probe both phases. To gain further insights one needs to disentangle the contributions of the different phases and to understand the details of dissociation and regeneration dynamics. Experimentally the path-length, lifetime and temperature dependence of the hadronic phase can be investigated via the system size or centrality dependence of  measurements in heavy-ion collisions. The larger the system (more central collisions) the longer the hadronic phase will last.

The EPOS3 model \cite{epos1,epos2,epos3} describes the full evolution of a heavy-ion collision. The initial stage is treated via a multiple-scattering approach based on Pomerons and strings. The reaction volume is divided into a core and a corona part \cite{epos4}. The core is taken as the initial condition for  the QGP evolution, for which one employ  viscous hydrodynamics. The corona part is simply composed of hadrons from string decays. After hadronisation of the fluid (core part), these hadrons and as well the corona hadrons are fed into UrQMD \cite{Bass:1998ca,Bleicher:1999xi}, which  describes hadronic interactions in a microscopic approach. The chemical and kinetic freeze-outs occur within this phase. The chemical freeze-out is expected to occur shortly after the phase transition from partonic to hadronic matter and is followed by the kinetic freeze-out.

In this paper we compare the resonance spectra and yields from EPOS with and without UrQMD to evaluate the main contributions from the later hadronic phase.  The hadronic phase might change the shapes of particle spectra (for stable particles as well as resonances) due to transverse flow, which builds up mostly in the hadronic phase.  Experimentally, resonances are reconstructed via their decay products, which are measured in a detector after leaving the hadronic medium.  Interactions in the hadronic medium can modify the yields of resonances, with both signal loss and signal gain possible.  Signal loss can occur if one or both of the decay products undergo pseudo-elastic scattering through a different resonance state, making it impossible to reconstruct the previous resonance state via the decay daughters due to the large change of the reconstructed invariant mass.  This signal loss is on the order of $10^{-1}$ of the total resonance yield.  In addition, elastic scattering of decay products can smear out the invariant-mass peak and cause further signal loss (on the order of $10^{-2}$).  These effects are most important in the low momentum region ($\pT\lesssim 2$~GeV/$c$) \cite{urqmd_highpt}.

However, signal gain is also possible: pseudo-elastic scattering through the inverse decay channel will regenerate a resonance and increase its measured yield.  This effect is also most important at low \pT and depends strongly on the lifetime of the resonance.  Recent results from the ALICE experiment show a suppression of the $\ks/K^{-}$ ratio in the most central \pb collisions with respect to elementary proton-proton collisions and with respect to the expected value from a thermal model description with a temperature of $T_{\rm ch}\sim 160$ MeV~\cite{alice_reso_pbpb}.  Similar suppression was not observed for the longer-lived \ph meson. This lead to the conclusion that interactions in the hadronic phase were likely responsible for the decrease in the reconstructible \ks resonance yields.

Thus, resonances are sensitive to the medium conditions throughout the whole expansion of the fireball. It is important to separate the different processes to obtain a better understanding of particle production and the contributions to the spectra from the partonic and hadronic phases.   In this paper we provide detailed information to determine the influence of the hadronic phase on the measured resonance signal.  We also investigate the mass and quark-flavor dependence of these effects. We show the centrality dependence of the spectra and yields of several resonances: \rh, \ks, \ph, \dl, \ssx, \ls, \xs, and their antiparticles. We compare the EPOS results for the \ks and the \ph to the measurements from the ALICE experiment~\cite{alice_reso_pbpb}. In addition, the $\ph/p$ ratio as a function of momentum showed a flattening with increasing centrality; the contributions to this effect of interactions in the hadronic phase will be investigated. These EPOS and UrQMD calculations focus on \pb collisions at \rsnn with centrality classes corresponding to those used by the ALICE experiment.

\section{EPOS model}

As explained in \cite{epos1,epos2,epos3,epos4}, EPOS3 is an event generator based on
3+1D viscous hydrodynamical evolution starting from flux tube initial
conditions, which are generated in the Gribov-Regge multiple scattering
framework. An individual scattering is referred to as a Pomeron, identified with a parton ladder, eventually showing up
as flux tubes (or strings). Each parton ladder is composed of a pQCD hard process, plus initial and final state linear parton emission.
Nonlinear effects are considered by using saturation scales $Q_{s}$, depending on the energy and the number of participants connected to the Pomeron in question.

The final state partonic system (corresponding to a Pomeron) amounts to (usually two) color flux tubes, being mainly longitudinal, with
transversely moving pieces carrying the \pT of the partons from hard scatterings. One has two flux tubes based on the cylindrical topology of the Pomerons. Each quark-antiquark pair in the parton ladder will cut a string into two; in this sense one may have more than two flux tubes. In any case, these flux tubes eventually constitute both bulk matter, also referred to as ``core'' \cite{eposcore} (which thermalizes, flows, and finally hadronizes) and jets (also referred to as ``corona''), according to some criteria based on the energy of the string segments and the local string density.
For the core, we use a 3+1D viscous hydrodynamic approach, employing a realistic equation of state, compatible with lQCD results.
We employ for all calculations in this paper a value of $\eta/s=0.08$.
Whenever a hadronisation temperature of $T_{H}$ is reached, we apply the usual
Cooper-Frye freeze-out procedure  \cite{cooperfrye}, to convert the fluid into particles.
We use $T_{H}=166\,$MeV. From this point on, we apply the hadronic
cascade UrQMD \cite{Bass:1998ca,Bleicher:1999xi}, about which more details are given later. All hadrons participate in the cascade, including those from the core (after freeze-out) and the corona. The corona particles, from string decay, are only ``visible" after a certain formation time (some constant of order one fm/$c$), multiplied by the corresponding gamma factor), so very high \pT particles have a good chance to escape.

\begin{figure}
\centering
\includegraphics[angle=270,scale=0.30]{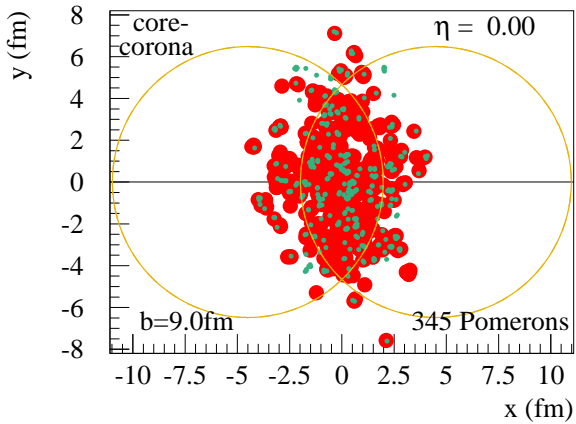}
\caption{Core-corona separation in a semi-peripheral (20-40\%) Pb-Pb
collision at \rsnn.}
\label{fig_core-example_1}
\end{figure}

\begin{figure}
\centering
\includegraphics[angle=270,scale=0.30]{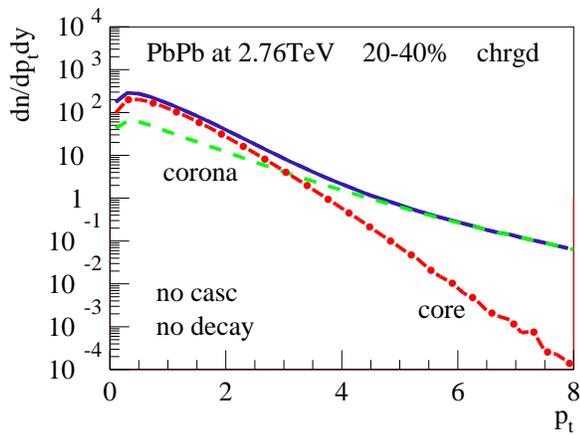}
\caption{Core and corona contributions
to the \pT spectra of charged hadrons in semi-peripheral (20-40\%) Pb-Pb
collisions at \rsnn.}
\label{fig_core-example_2}
\end{figure}

\begin{figure}
\centering
\includegraphics[angle=270,scale=0.30]{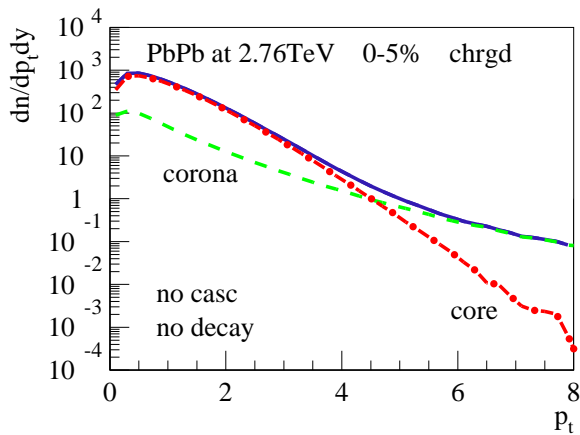}
\caption{Core and corona contributions
to the \pT spectra of charged hadrons in central Pb-Pb
collisions.}
\label{fig_core-example_3}
\end{figure}

EPOS3 is a universal approach, in the sense that it employs the same approach for $pp$, $pA$, and $AA$ scatterings, with the same core-corona procedure. In figs.~\ref{fig_core-example_1},\ref{fig_core-example_2}~and~\ref{fig_core-example_3},  we discuss an
example of core-corona separation in a semi-peripheral (20-40\%) Pb-Pb collision at \rsnn. Figure~\ref{fig_core-example_1} shows string segments in the transverse plane from the core (red) and corona (green). Even in a small system (peripheral collisions) there are sufficient overlapping core string segments to provide a
core of plasma matter, allowing a (short) hydrodynamic expansion, which quickly
builds up flow. In figure~\ref{fig_core-example_2}, we plot the contribution from
core and corona to the \pT spectra of charged hadrons in the 20-40\% most central Pb-Pb collisions. Figure \ref{fig_core-example_3} shows the core and corona contributions to the \pT spectra of changed hadrons in central collisions Pb-Pb, where the particle contribution from the core is increased with respect to the corona contribution. This leads into an increase of the charged hadrons in the low momentum range coming form the core contribution, which mainly defines the interacting hadronic medium till the kinetic freeze-out. We also observe an increase of the core contribution out to higher momenta.

%% maybe
%\subsection{Core and Corona of the EPOS model}
%\begin{figure}[h!]
%\centering
%\includegraphics[scale=0.45]{plots/core_corona.eps}
% \caption{Momentum distribution of changed particles at mid-rapidity coming from the core and corona part} 
% \label{core_corona}
%\end{figure}
%
%

% \subsection{Centrality definition} 

%\vspace{2cm}

 \section{UrQMD} 

The UrQMD model is a non-equilibrium transport approach. The interactions of hadrons in the current version include binary elastic and $2 \rightarrow n$ inelastic scatterings, resonance creations and decays, string excitations, particle + antiparticle annihilations as well as strangeness exchange reactions \cite{Graef:2014mra}. The cross sections and branching ratios for the corresponding interactions are taken from experimental measurements (where available), detailed balance relations and the additive quark model. The model describes the full phase-space evolution of all hadrons, including resonances, in a heavy-ion collision based on their hadronic interactions and their decay products. Due to the short lifetime of resonances, their decay products may interact in the hadronic phase. This is not the case for weak decays, where the system has already decoupled at the time of the decay. As discussed previously, the experimental reconstruction of resonances will be influenced by the final state interactions of the decay products. Resonance signals have been previously studied using the UrQMD model~\cite{urqmd_high_pt,urqmd_sis,urqmd_2002,urqmd_2003,urqmd_2004,urqmd_2005,urqmd_2006,urqmd_2008,Steinheimer:2015msa,Steinheimer2012}.
This paper will provide a detailed study of the survival of resonances which are measurable, depending on momentum, lifetime, and decay time, as well as the collision centrality.

\section{Resonance Reconstruction Data and Model}

Experimentally, hadronic resonances are mainly reconstructed via the momentum of their charged decay daughters, which are identified trough the measurements of energy loss $(dE/dx)$ in a Time Projection Chamber (TPC) and/or the velocity in a Time of Flight (TOF) detector. Weakly decaying particles, such as $\Lambda$ and $\Xi$ can be selected topologically via their decay vertex position, which adds an additional constraint. 
The investigated hadronic decays within the EPOS3 approach are listed in table~\ref{decay}~\cite{PDG}.
Quoted reconstructable resonance yields are investigated in these specific decay channels, which are the same channels used experimentally, and corrected using the corresponding branching ratio. In these model calculations, resonances that decay by the channels listed in Table I are flagged and the decay products are followed throughout the system evolution.  If neither decay product undergoes a re-scattering, the resonance is flagged as reconstructable.

\begin{table}[h!]
\centering
\caption{The resonances are constructed experimentally via their listed decay channels~\cite{PDG}. These decays are used in EPOS and UrQMD for the resonance calculations.}
\label{table2}
\begin{ruledtabular}
\begin{tabular}{llccc}
Resonance & decay channel & branching ratio & lifetime (fm/$c$) \\
\hline
$\rho(770)^{0}$ & $\pi^{+}$ + $\pi^{-}$  & 1 & 1.335 \\
$K^{*}(892)^{0}$ & $\pi^{-}$ + $K^{+}$ & 0.67 & 4.16 \\
$\phi(1020)$ & $K^{+}$ + $K^{-}$ & 0.489 & 46.26 \\
$\Delta(1232)^{++}$ & $\pi^{+}$ + $p$ & 1 & 1.69 \\
$\Sigma(1385)^{+}$ & $\pi^{+}$ + $\Lambda$ & 0.870 & 5.48 \\
$\Sigma(1385)^{-}$ & $\pi^{-}$ + $\Lambda$ & 0.870 & 5.01 \\
$\Lambda(1520)$ & $K^{-}$ + $p$ & 0.225 & 12.54\\
$\Xi(1530)^{0}$ & $\pi^{+}$ + $\Xi^{-}$ & 0.67 & 22 \\
\end{tabular}
\end{ruledtabular}
\label{decay}
\end{table}

\section{Centrality}

In the ALICE Experiment, centrality determination for \pb collisions at \rsnn is done via the measurement of the charged-particle multiplicity using the VZERO detectors \cite{centrality1,centrality}, which span the pseudorapidity ranges $-3.7<\eta<-1.7$ and $2.8<\eta<5.1$. The impact parameter and geometrical quantities, such as number of participants ($N_{\rm part}$), number of spectators ($N_{\rm spec}$), or number of collisions ($N_{\rm coll}$), are not directly measurable. The experimental observable related to the collision geometry is the average charged-particle multiplicity ($N_{\rm ch}$), which decreases monotonically with increasing impact parameter.

\begin{figure}
\centering
\includegraphics[scale=0.40]{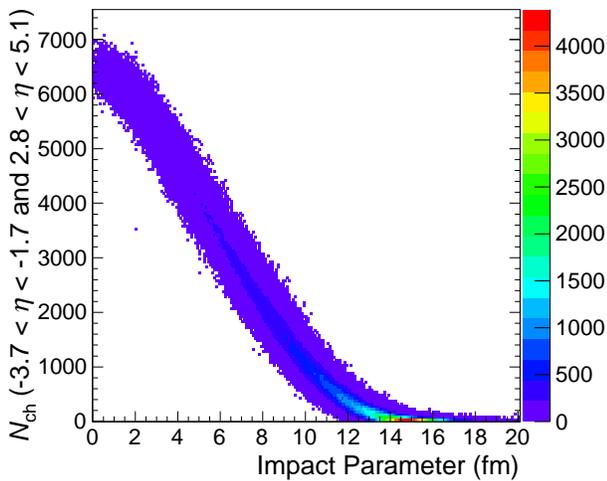}
 \caption{EPOS3 calculation of the charged-particle multiplicity inside the acceptance of the ALICE VZERO detector ($-3.7<\eta<-1.7$ and $2.8<\eta<5.1$) as a function of impact parameter.}
 \label{impact}
\end{figure}

Figure~\ref{impact} shows the charged-particle multiplicity distribution as a function of impact parameter in EPOS in the acceptance region of the two VZERO detector systems. The centrality intervals for the EPOS results in this paper are determined directly via the simulated collision impact parameter.  The boundaries of the various centrality intervals used for EPOS3 agree with the values calculated by the ALICE Collaboration~\cite{ALICE_multiplicity} within 1\%. For this study we generated about 800 thousand collisions using EPOS3 with UrQMD turned on and 450 thousand events using EPOS3 with UrQMD turned off.

 \section{Spectra}  
 
\begin{figure}
\centering
\includegraphics[scale=0.45]{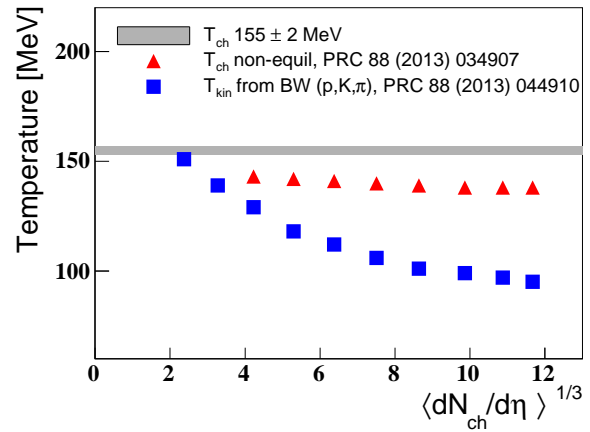}
\vspace{-0.8cm}
\caption{Centrality dependence of the chemical and kinetic freeze-out temperatures of the hadronic phase \cite{thermal_raf,blastwave_pbpb} as a function of centrality via \dncr. } 
\label{temp_chem_kin}
\end{figure}

Comparison of the resonance results from EPOS3 with and without the UrQMD phase allows the changes in the spectra and yields due to the hadronic phase to be investigated. From blast-wave fits, one expects in central collisions the longest lifetime of the hadronic phase and the lowest kinetic freeze-out temperature of $T_{\rm kin} = 90$~MeV \cite{blastwave_pbpb}. Figure~\ref{temp_chem_kin} shows the centrality dependence of the chemical and kinetic freeze-out temperatures for 2.76 TeV Pb-Pb collisions, in the blast-wave approach.
%%
%%%%%%%%%%%%%%%%%%%%%%%%%%%%%%%%%%%%%%%%%%%%%%%%%%%%%%%%%%%%%%%%%%%%%%%%%%%%%%%%%%%%%%%%%%%%%%%%%%%%%%%%%%%%%%%%%%%
%{\color{red}   REMARK: Could be quite different in EPOS. Would be good to have some information (about lifetime of hadronic phase), as already discussed. Some conclusions in the following are based on the blastwave  picture, whereas EPOS is a bit more complicated due to the core-corona picture, which is centrality dependent } 
%%%%%%%%%%%%%%%%%%%%%%%%%%%%%%%%%%%%%%%%%%%%%%%%%%%%%%%%%%%%%%%%%%%%%%%%%%%%%%%%%%%%%%%%%%%%%%%%%%%%%%%%%%%%%%%%%%%%
%%
%(xxxplot Tch and Tkin )?
\vspace{-0.5cm}

\begin{figure}
\centering
\includegraphics[scale=0.35]{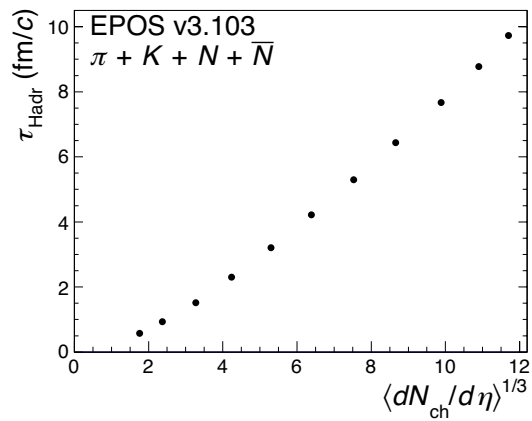}
\vspace{-0.3cm}
\caption{Centrality dependence of the lifetime of the hadronic phase calculated from the estimated difference in the hadronic-phase lifetime between the EPOS+UrQMD ON and EPOS+UrQMD OFF scenarios, calculated using hadrons ($\pi,K,N$ and $\overline{N}$).}

%The values of the lifetime difference are plotted as a function of the values of \dncr.}

%measured by the ALICE experiment~\cite{ALICE_multiplicity} at mid rapidity ($|\eta|<0.5$).  For the 90-100\% centrality interval (where no measured \dnc values have been reported), the EPOS3 data are plotted at the \dncr values found in these simulations.}
\label{epos_lifetime}
\end{figure}

\begin{figure*}
\centering
\includegraphics[scale=0.80]{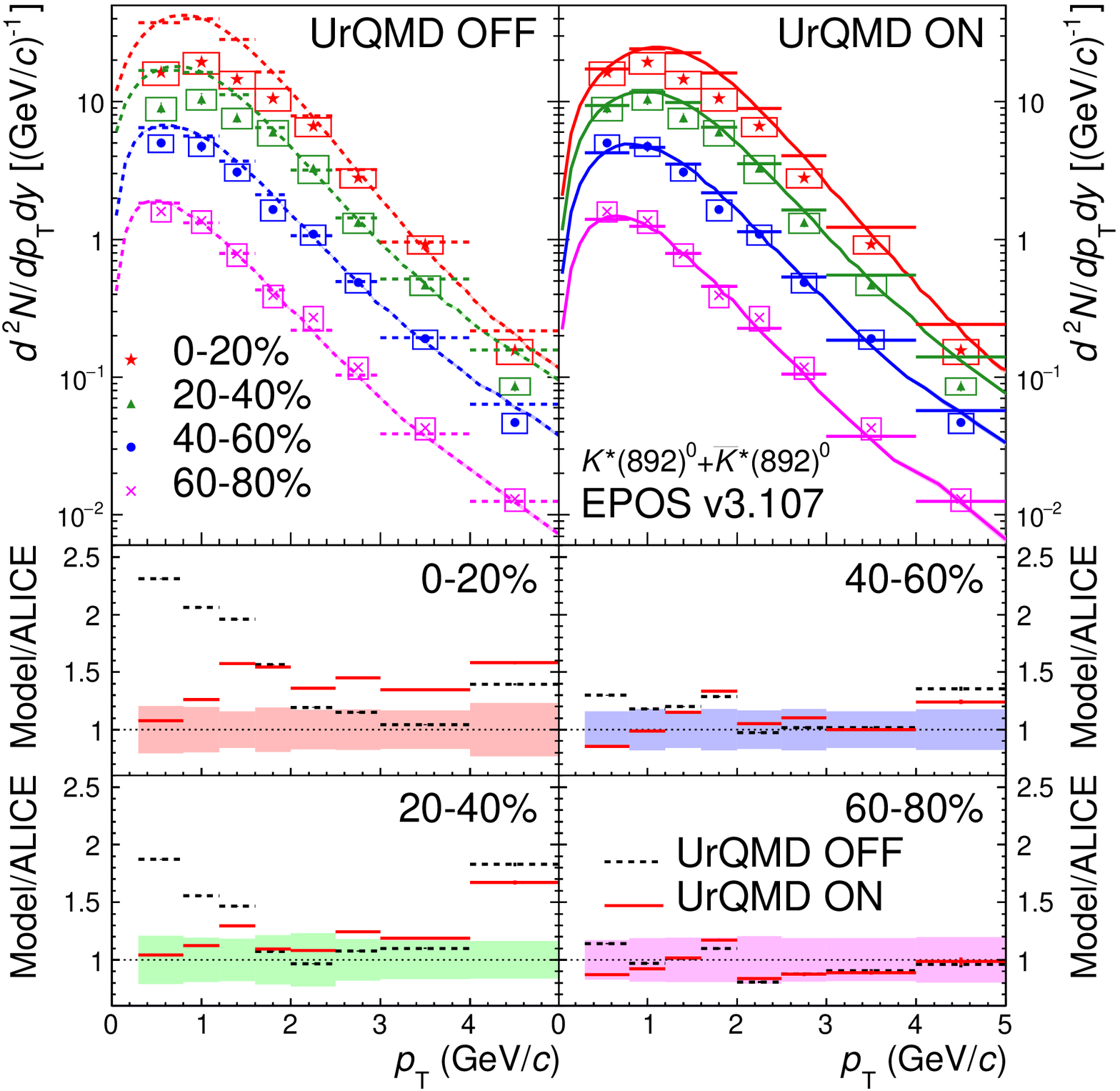}[b]
\caption{\textbf{Upper panels:} transverse momentum distributions of $\ks+\aks$ from EPOS3 with UrQMD OFF and UrQMD ON for \pb collisions at \rsnn in different centrality intervals.  These results are compared to measurements from the ALICE Experiment~\cite{alice_reso_pbpb} (same data on left and right).  The curves are the EPOS3 \pT distributions with fine bins and the shaded bands are the statistical uncertainties of the EPOS3 results.  The horizontal lines are the EPOS3 results in the same bins as the ALICE measurements.  \textbf{Lower panels:} the ratio of the EPOS3 results to the ALICE measurements as functions of \pT for the different centrality intervals.  The shaded bands around unity represent the uncertainties of the measured data.} 
\label{PbPb_Kstar0_spectra_UrQMDX}
\end{figure*}

Figure~\ref{epos_lifetime} shows the lifetime difference between EPOS with the UrQMD hadronic phase (EPOS+UrQMD ON) and 
EPOS without the UrQMD hadronic phase (EPOS+UrQMD OFF), which defines the lifetime of the interactions in the hadronic phase. 
The estimated lifetime of the hadronic phase is the difference of the production time of stable hadrons ($\pi,K,N$ and $\overline{N}$) between the EPOS+UrQMD ON $\langle \tau_{on}\rangle$ and EPOS+UrQMD OFF $\langle \tau_{off}\rangle$ scenarios: \\
$ \tau_{Hadr} $ = $\langle \tau_{on}\rangle$  - $\langle \tau_{off}\rangle$

\begin{figure*}
\centering
\includegraphics[scale=0.80]{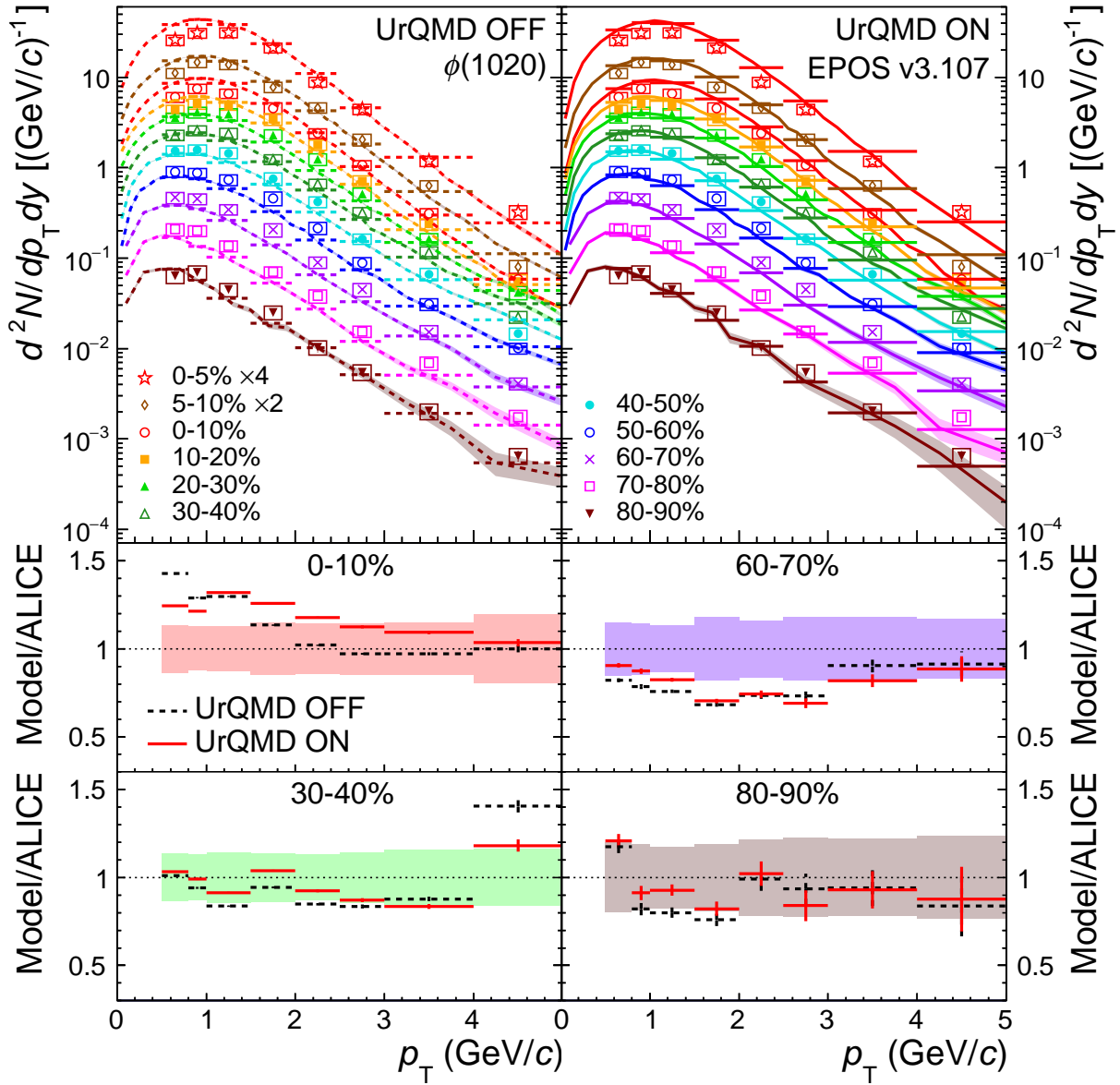}
\caption{\textbf{Upper panels:} transverse momentum distributions of \ph from EPOS3 with UrQMD OFF and UrQMD ON for \pb collisions at \rsnn in different centrality intervals.  These results are compared to measurements from the ALICE Experiment~\cite{alice_reso_pbpb} (same data on left and right).  The curves are the EPOS3 \pT distributions with fine bins and the shaded bands are the statistical uncertainties of the EPOS3 results. The horizontal lines are the EPOS3 results in the same bins as the ALICE measurements.  \textbf{Lower panels:} the ratio of the EPOS3 results to the ALICE measurements as functions of \pT for four selected centrality intervals.  The shaded bands around unity represent the uncertainties of the measured data.} 
\label{PbPb_phi_spectra_UrQMDX}
\end{figure*}

We study the centrality dependence of the resonance spectra to explore the lifetime and the associated kinetic freeze-out temperature dependence of the hadronic phase. The spectra of the \ks and the \ph from EPOS3 without and with UrQMD and their ratios to data as functions of centrality are shown in figures~\ref{PbPb_Kstar0_spectra_UrQMDX} and \ref{PbPb_phi_spectra_UrQMDX}.

\begin{figure*}
\centering
\begin{minipage}{0.9\textwidth}
\includegraphics[scale=0.80]{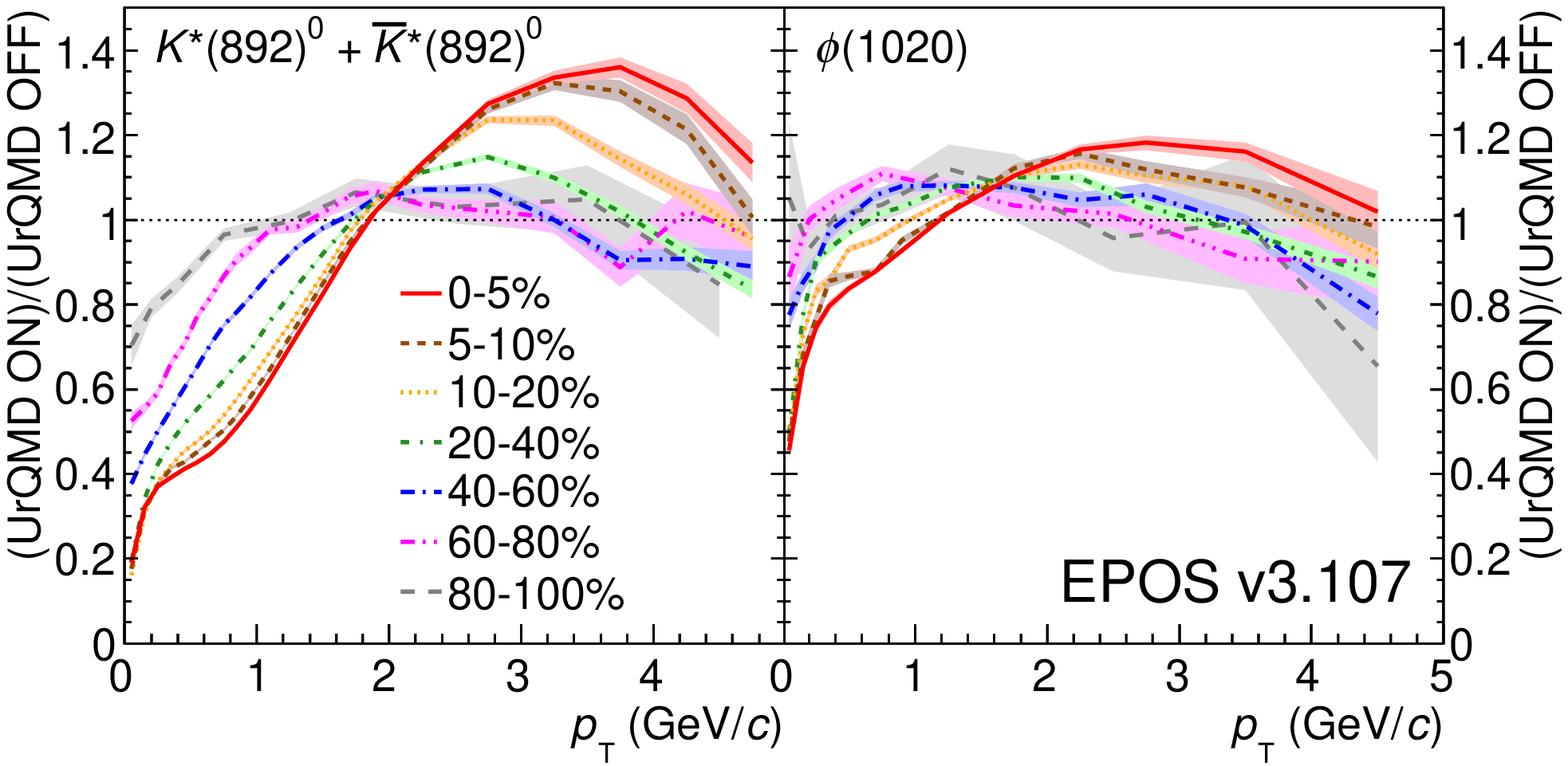}
 \caption{The effect of the UrQMD phase on the \pT spectra of the \ks and \ph mesons is illustrated by this \pT-dependent ratio: the \pT spectra from EPOS3 with UrQMD ON divided by the spectra with UrQMD OFF in different centrality intervals.  The shaded bands represent the statistical uncertainties of the ratio.} 
\end{minipage}
\label{UrQMD_ratio_figure}
\end{figure*}

Figure~\ref{PbPb_Kstar0_spectra_UrQMDX} shows the influence of the hadronic phase on the \ks via the spectra from EPOS3 without UrQMD (``UrQMD OFF") (left) and EPOS3 with UrQMD (``UrQMD ON") (right). The shaded area indicates the statistical and systematic uncertainties of the measured data. Clearly EPOS3 with UrQMD ON allows for a better description of the \ks spectra than EPOS3 with UrQMD OFF. The suppression of the \ks signal in the low-\pT region of the 0-20\% and 20-40\% centrality intervals is described by re-scattering in the hadronic phase.  This effect is larger the longer the hadronic phase lives, \textit{i.e.,} in the most central collisions (0-20\%). In  peripheral collisions (40-60\% and 60-80\%), where the hadronic lifetime is short, both spectra, URQMD OFF and ON, are very similar.

Since the lifetime of the \ph is long ($\tau$~=~46~fm/$c$)~\cite{PDG} compared to the fireball lifetime ($\tau\approx$ 10-15~fm/$c$) \cite{Lin,Graef2012} most of these resonances decay after the hadronic phase. Therefore a maximum signal loss of only $\sim$10\% in the most central collisions is expected. Figure~\ref{PbPb_phi_spectra_UrQMDX} shows the agreement with the ALICE measurements of the EPOS3 with UrQMD OFF and ON scenarios. This indicates that measurements of the \ph from its hadronic decay channel are not sensitive to interactions in the hadronic phase. However, in the most central collisions we observe a small signal loss in the low-\pT region, which can be explained by a small re-scattering effect. The initial \ph yield in EPOS3 seems a little too large in the low-\pT region. \\

The hadronic phase has a larger effect on the \ks than the \ph, shown as a difference between the URQMD ON and URQMD OFF spectra and as the ratio (URQMD ON)/(URQMD OFF) in figure~9.  %xxx\ref{UrQMD_ratio_figure}. 
The \ks is suppressed by at least 50\% in the low-\pT region $\pT<1$~\gvc and the \ph exhibits smaller low-\pT suppression. The largest modification to the spectra is for both particles is for central collisions. This will result in a change of the mean transverse momentum.

\section{Mean Transverse Momentum}

\begin{figure}
\includegraphics[scale=0.42]{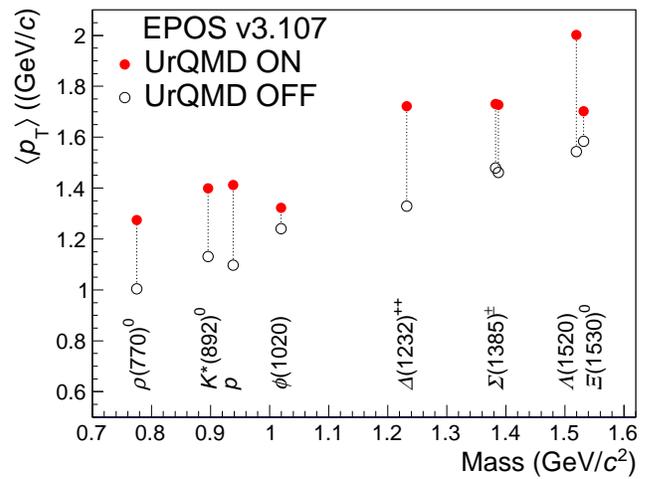}
 \caption{Mean transverse momentum (\mpt) versus hadron mass from EPOS3 with UrQMD off and UrQMD on for central (0-10\%) \pb collisions at \rsnn.}
\label{mpt_all}
\end{figure}

Figure~\ref{mpt_all} shows the change of the mean transverse momentum (\mpt) in central \pb collisions at \rsnn as a function of hadron mass due to interactions in the hadronic phase. Overall an increase in \mpt of 200-300 MeV/c is visible for most of the resonances and the proton. Only a small increase in \mpt is observed for the \ph and \xs due to the lower interaction rate in the hadronic medium of the daughter particles due to the comparatively long lifetimes of those two particles. In addition a larger regeneration for the \xs from the $\Xi$+$\pi$ channel is expected which compensates the signal loss in the low momentum region. \\

% and/or a compensating regeneration within in the same momentum range as re-scattering (for \xs) .

\begin{figure*}
\centering
\includegraphics[scale=0.85]{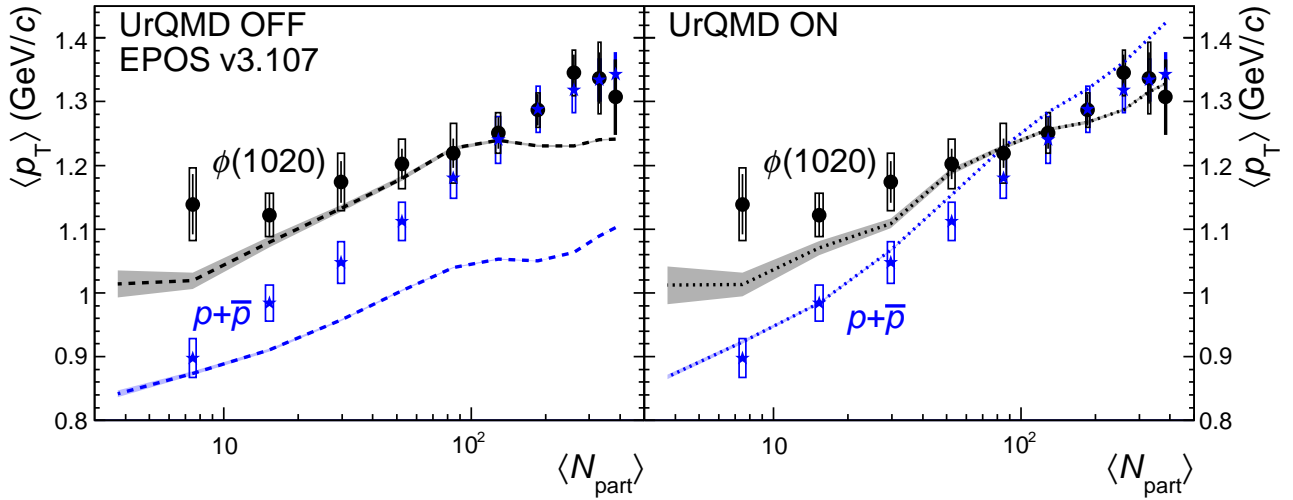}
 \caption{Mean transverse momentum \mpt of (anti)protons and \ph given by EPOS3 for the UrQMD OFF (left panel) and UrQMD ON (right panel) scenarios for multiple centrality intervals.  Shaded bands represent the statistical uncertainties.  These results are compared to measurements from the ALICE Experiment~\cite{alice_reso_pbpb} (same data in both panels); statistical and systematic uncertainties are represented by bars and boxes, respectively.  The \mpt values (both theoretical and experimental) are plotted as functions of the mean number of participant nucleons for each centrality class found by the ALICE Collaboration~\cite{centrality}.}
\label{mpt_p_phi}
\end{figure*}

The ALICE Collaboration observed that (anti)protons and \ph mesons have very similar mean transverse momentum (\mpt) values for central and mid-central \pb collisions, but observed a splitting in the \mpt values for these particles for peripheral collisions, with \ph mesons having greater \mpt than (anti)protons~\cite{alice_reso_pbpb}.  EPOS3 results for the \mpt of these particles are compared to the ALICE measurements in Fig.~\ref{mpt_p_phi}.  The (anti)proton \mpt values are well described, though slightly overestimated, by EPOS3 with UrQMD ON, but not well described by EPOS3 without UrQMD. This means that the elastic interactions and feed-down from resonance decays in the hadronic phase are needed to describe the proton momentum spectrum. The \ph momentum spectrum itself is negligibly affected by the hadronic phase. Only a small increase in the mean transverse momentum in very central collisions due to directed flow is visible in central collisions (Fig.~\ref{mpt_p_phi} (right)).

\begin{figure*}
\centering
\includegraphics[scale=0.43]{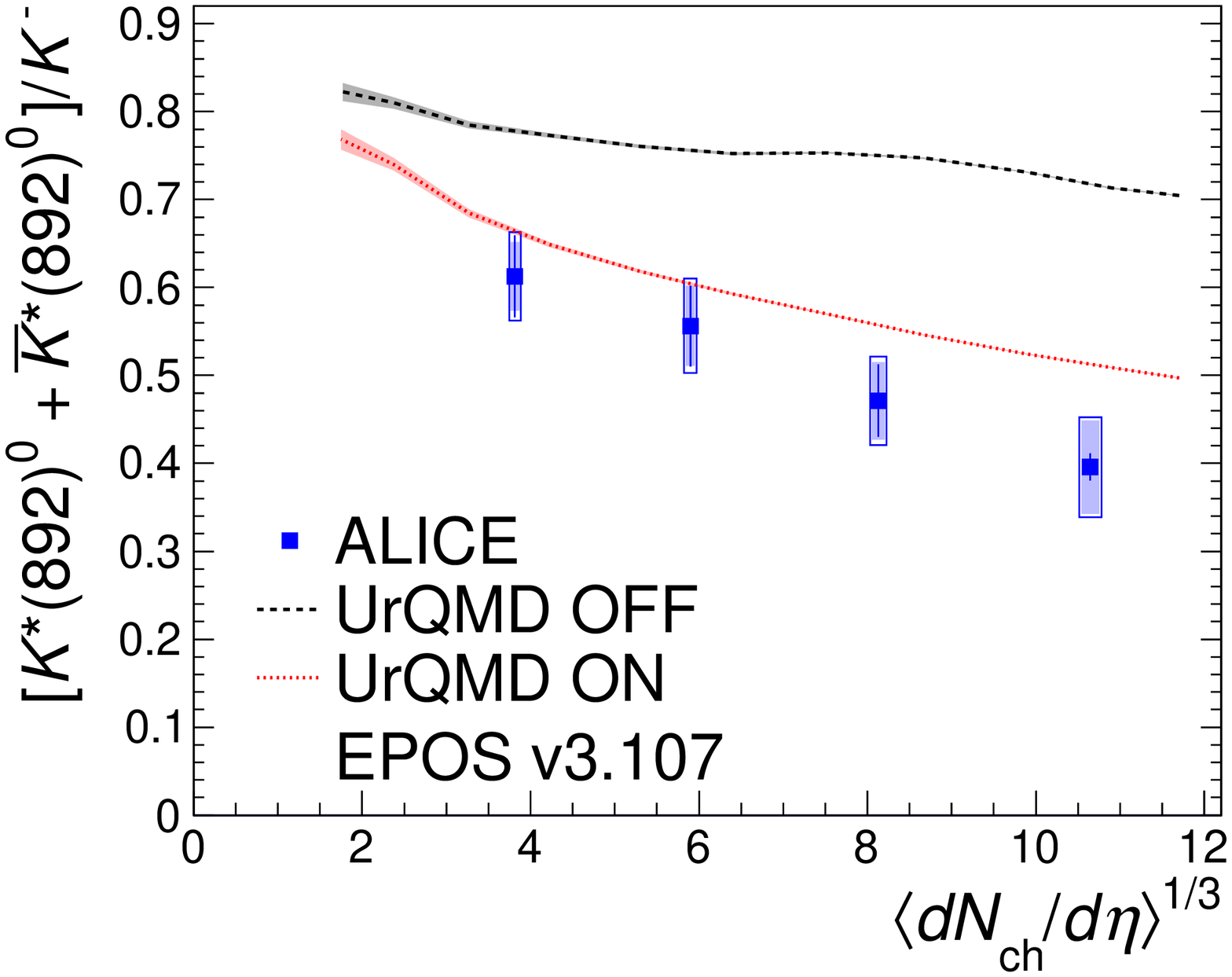}
\includegraphics[scale=0.43]{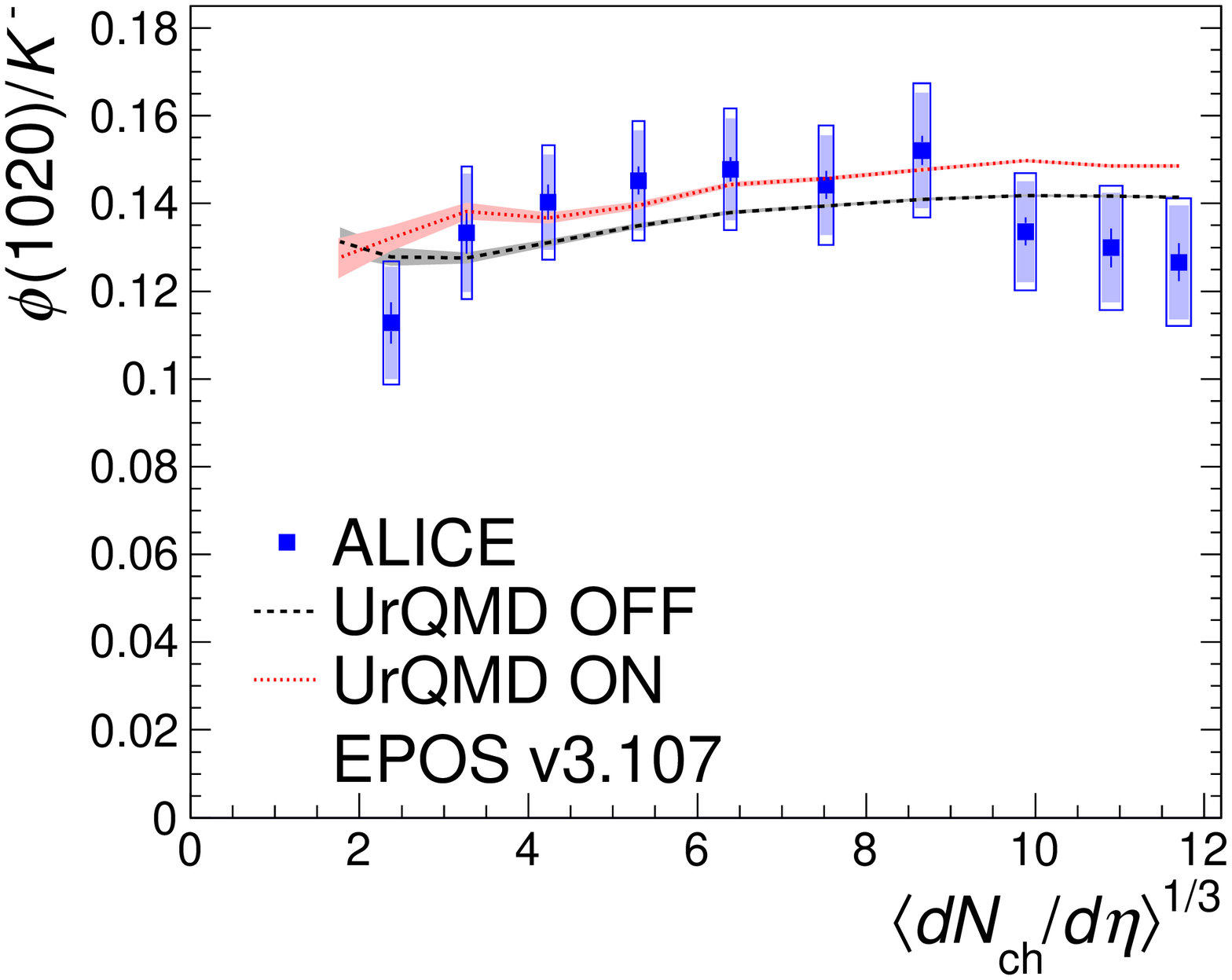}
 \caption{Ratio of integrated yields $(\ks+\aks)/K^{-}$ (left) and $\ph/K^{-}$ (right) for multiple centrality intervals calculated using EPOS3 with UrQMD OFF and UrQMD ON compared to measurements by the ALICE Experiment~\cite{alice_reso_pbpb}.  The shaded bands around the EPOS3 curves represent their statistical uncertainties.  The bars on the ALICE measurements show the statistical uncertainties.  The shaded boxes represent the systematic uncertainties that are uncorrelated between centrality intervals, while the open boxes represent the total systematic uncertainties.  The experimental and theoretical data are both plotted as functions of the values of \dncr measured by the ALICE experiment~\cite{ALICE_multiplicity} at mid rapidity ($|\eta|<0.5$).  For the 90-100\% centrality interval (where no measured \dnc values have been reported), the EPOS3 data are plotted at the \dncr values found in these simulations.} 
\label{Kstar2Km_phi2Km}
\end{figure*}

\begin{figure*}
\centering
\includegraphics[scale=0.88]{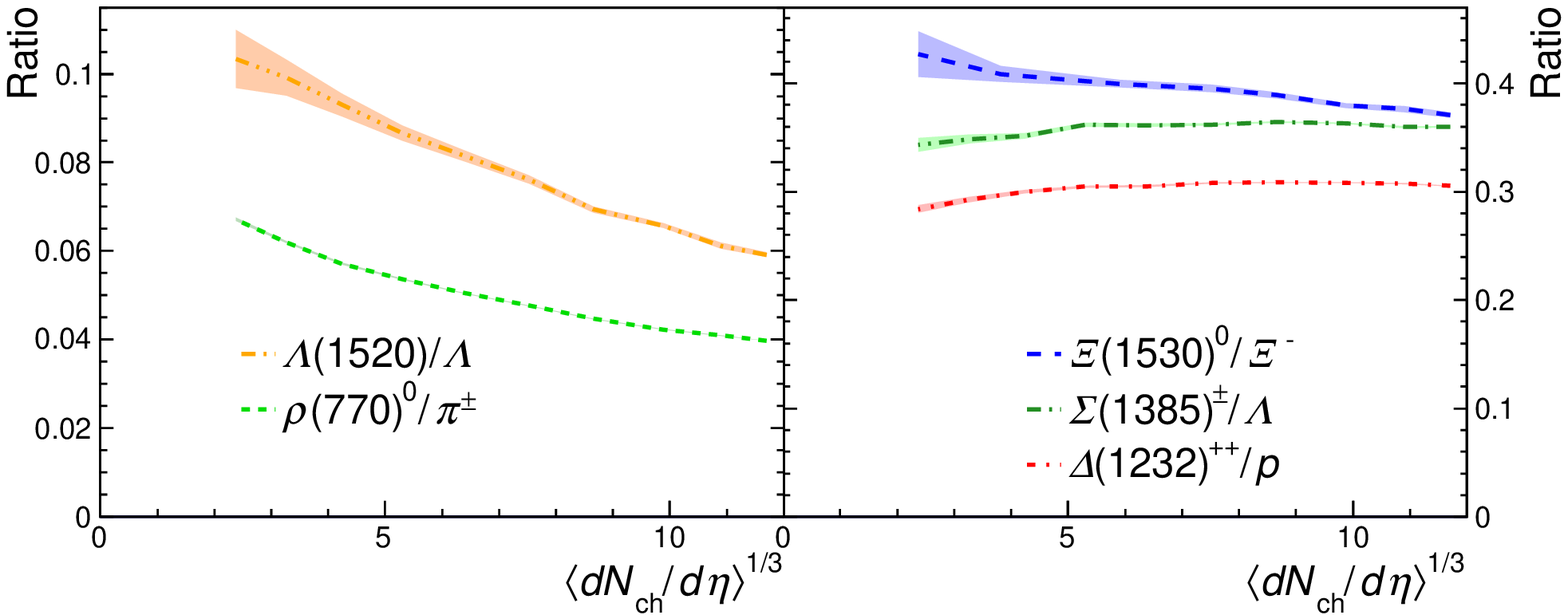}
 \caption{Ratio of integrated yields $\ssx/\Lambda$  and $\rh/\pi^{\pm}$ (left) $\xs/\Xi^{-}$ ,  $\ssx/\Lambda$ and  $\dl/p$ (right) for multiple centrality intervals calculated using EPOS3 with UrQMD ON (numerators and denominators are sums of particles and antiparticles).  The shaded bands around the EPOS3 curves represent their statistical uncertainties.  The theoretical data are plotted as functions of the values of \dncr measured by the ALICE experiment~\cite{ALICE_multiplicity} at mid rapidity ($|\eta|<0.5$). }
 \label{nonmeasured_ratios}
\end{figure*}

\section{Yields and Ratios}

Fig.~\ref{Kstar2Km_phi2Km} shows the $\ks/K^{-}$ and $\ph/K^{-}$ ratios calculated using EPOS3 with UrQMD OFF and UrQMD ON for various centrality intervals.  (In this context, the symbol \ks refers to the sum of the \ks and \aks.)  These are compared to the values measured by the ALICE Experiment~\cite{alice_reso_pbpb}.  The ratios (both experimental and theoretical) are shown as functions of the cube root of the mean mid-rapidity ($|\eta|<0.5$) charged-particle multiplicity measured by the ALICE Experiment~\cite{ALICE_multiplicity}.  This abscissa is used because femtoscopy studies~\cite{ALICE_HBT_2011,Lisa_FemtoscopyReview,Graef2012} suggest that it scales in proportion to the radius of the collision system.  Under the simple assumption that the probability of re-scattering is proportional to the distance traveled through the hadronic medium, an exponential decrease in measured resonance yields as a function of the system radius or \dncr might be expected.  The measured $\ks/K^{-}$ ratio exhibits suppression, which increases in strength with collision centrality.  In contrast, the measured $\ph/K^{-}$ has only a weak dependence on centrality and is not suppressed in central collisions.  The EPOS3 model with UrQMD ON provides a fair description of the $\ks/K^{-}$ ratio: it reproduces the decreasing behavior of the ratio as a function of centrality, but gives less suppression in central collisions than was observed. EPOS3 with UrQMD OFF overestimates the value of the $\ks/K^{-}$ ratio and predicts only a weak centrality dependence. The $\ph/K^{-}$ ratio is reproduced well by EPOS3 with both UrQMD ON and OFF. This confirms the explanation that the hadronic phase does not change the yield of reconstructable \ph very much due to its long lifetime, which would cause most of the decays ($\sim$90\%) to occur after the kinetic freeze-out.

Figure~\ref{nonmeasured_ratios} shows the $\rh/\pi^{\pm}$, $\dl/p$, $\ssx/\Lambda$, $\ls/\Lambda$, and $\xs/\Xi^{-}$ ratios predicted for various centrality intervals by EPOS3 with UrQMD ON.  (We always plot the ratios of the sums of particles and antiparticles, but leave the antiparticles out of the text for brevity.)  As before, the ratios are presented as a function of the experimental \dncr values~\cite{ALICE_multiplicity}.  Strong suppression is expected in the $\rh/\pi^{\pm}$ and $\ls/\Lambda$ ratios, with increasing suppression for central collisions. The same qualitative behavior is observed for the $\ks/K^{-}$ ratio. 
%% ch new
It is noticeable that the lifetime of the \ls is about three times larger than the \ks lifetime. Therefore more \ks are decaying during the hadronic phase and therefore more decay particles are affected by hadronic re-scattering.  Since the suppression for the \ks is not larger than for the \ls a larger regeneration cross section for the \ks than for the \ls is expected. This trend has been confirmed in measurements at RHIC energies \cite{markert1}. 
%%%
In contrast, the $\dl/p$, $\ssx/\Lambda$, and $\xs/\Xi^{-}$ ratios do not have large centrality dependences. A possible explanation would be large regeneration cross-sections for these baryonic resonances. In addition, the \xs has a comparatively long lifetime (about 5 times the lifetime of \ks) and may be less sensitive to hadronic re-scattering effects, as was the case for \ph. For the short lived $\ssx$ we expect a large regeneration cross section via the $\Lambda+\pi$ channel. This result is also in agreement with the centrality-independent $\ssx/\Lambda$ ratio, which has been observed for lower energy collisions at RHIC \cite{markert1}.

\begin{figure*}
\centering
\includegraphics[scale=0.88]{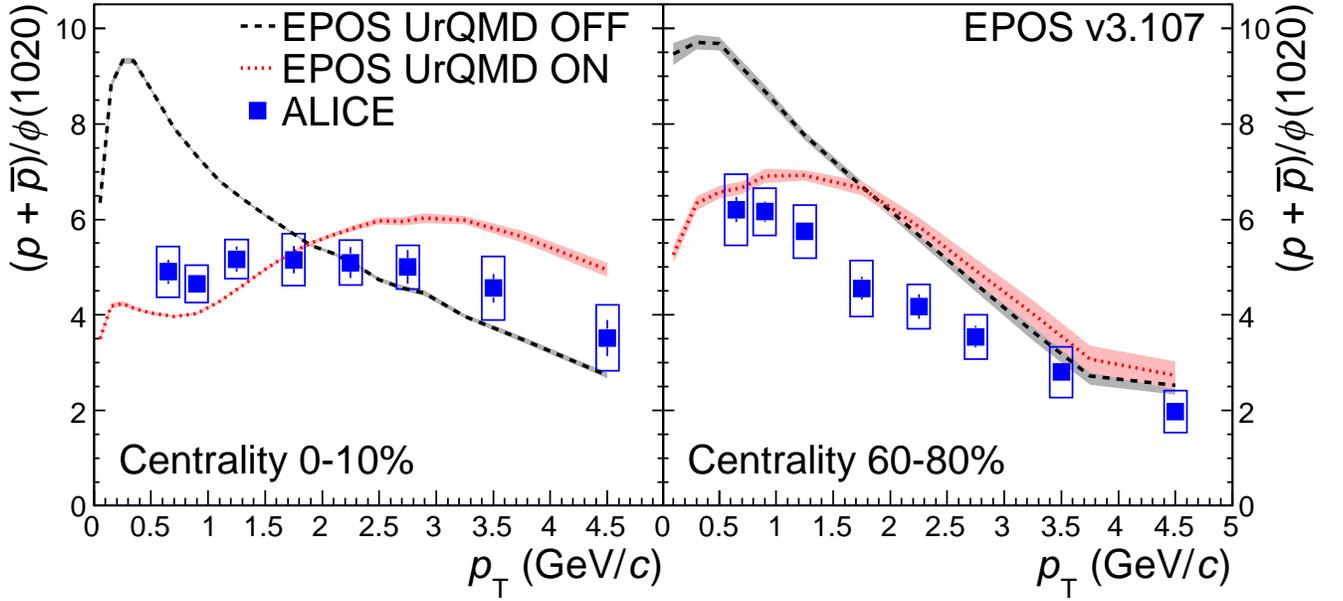}
 \caption{The \pT-dependent (anti)proton/\ph ratio for the 0-10\% (left) and 60-80\% (right) centrality intervals, calculated using EPOS3 and measured by the ALICE Experiment~\cite{alice_reso_pbpb}.  For the EPOS3 curves the shaded bands represent the statistical uncertainties.  For the ALICE data bars and boxes represent statistical and systematic uncertainties, respectively.}
\label{p2phi}
\end{figure*}

In Ref.~\cite{alice_reso_pbpb} it was shown that the (anti)proton/\ph ratio is independent of \pT for $\pT<4$~\gvc in central \pb collisions (consistent with the expectations from hydrodynamic models for particles with similar masses), but becomes sloped for peripheral collisions.  As shown in Fig.~\ref{p2phi} EPOS3 + UrQMD is able to reproduce this behavior.  For central collisions, EPOS3 with UrQMD OFF does not describe the \pT-dependent (anti)proton/\ph ratio.

\begin{figure}
\centering
\includegraphics[scale=0.43]{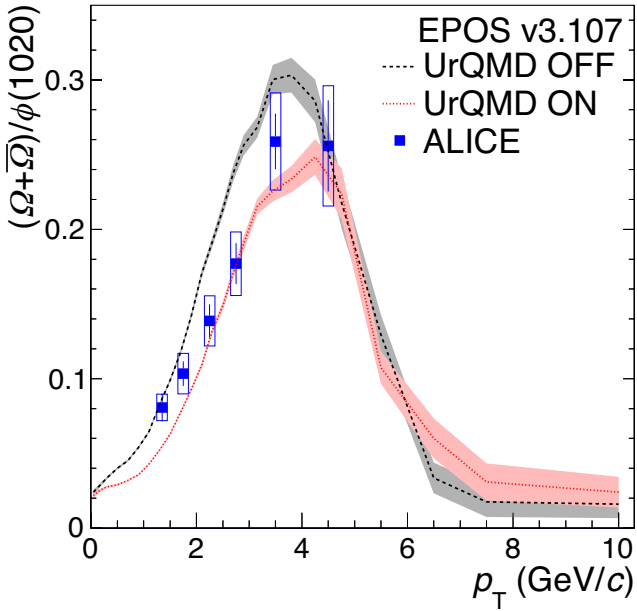}
		 \caption{The \pT-dependent ($\Omega+\bar{\Omega}$)/\ph ratio for the 0-10\% centrality intervals, calculated using EPOS3 and measured by the ALICE Experiment~\cite{alice_omega,alice_reso_pbpb}.  For the EPOS3 curves the shaded bands represent the statistical uncertainties.  For the ALICE data bars and boxes represent statistical and systematic uncertainties, respectively.}
\label{omega2phi}
\end{figure}

EPOS3 with UrQMD ON gives a fair description of the measured data, though it does exhibit some \pT dependence that is not seen in the measured data.  For peripheral collisions, EPOS3 with UrQMD ON gives a qualitative description of the measured data, but slightly overestimates the values of the ratio for most \pT bins.

The hadronic interaction cross sections of the \ph and $\Omega$ are very small. In addition there is not expected to be a large contribution from feed-down from higher mass resonances. Therefore the ($\Omega+\bar{\Omega}$)/\ph ratio from EPOS with and without the UrQMD phase for central collisions (0-10\%) is very similar as shown in figure~\ref{omega2phi}. The EPOS3 model describes the ($\Omega+\bar{\Omega}$)/\ph ratio measured by the ALICE experiment very well \cite{alice_omega,alice_reso_pbpb}. Therefore the $\Omega$ and $\phi$ are sensitive to the conditions of the early system, and the $\Omega/\phi$ ratio serves as good probe for the directed flow at hadronization.

\section{Appendix: Transverse Momentum Spectra Predictions for other Resonances}

\begin{figure*}
\centering
\includegraphics[scale=0.8]{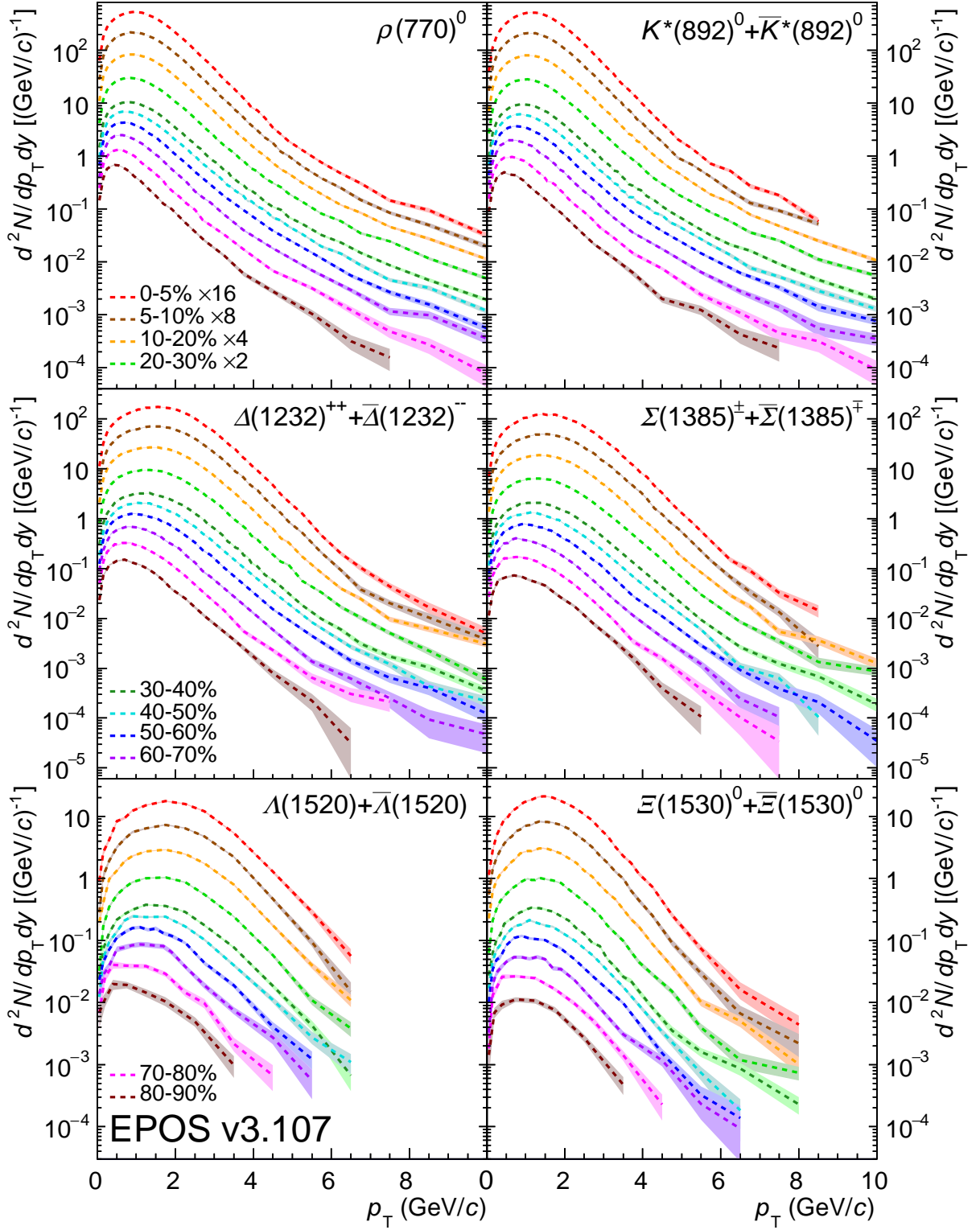}
\caption{Transverse-momentum spectra from EPOS3 with UrQMD ON (right panels) for the \rh, \dl, \ssx, \ls, \xs (summed with their antiparticles) in different centrality intervals. The shaded bands represent the statistical uncertainties.} 
\label{nonmeasured_spectra_UrQMD}
\end{figure*}

Fig.~\ref{nonmeasured_spectra_UrQMD} shows predicted \pT distributions from EPOS3 with UrQMD ON for \rh, \dl, \ssx, \ls, and \xs (summed with their antiparticles) in multiple centrality intervals (as well as for the $\ks+\aks$ in finer centrality intervals than reported in Fig.~\ref{PbPb_Kstar0_spectra_UrQMDX}). The magnitude of the re-scattering effect is predominantly given by the lifetime of the decay.

\section{Summary}

EPOS3 using UrQMD as hadronic afterburner approach is well suited to describe the experimentally measurable resonances in heavy ion reactions at LHC energies. 
The increasing suppression of the $\ks/K^{-}$ ratio towards more central Pb-Pb collisions is explained by interactions in the
extended hadronic phase until the system reaches the kinetic freeze-out. This suppression is not seen in the \ph meson due to its longer lifetime. This observation is in agreement with the EPOS3 predictions in case of UrQMD turned on. Furthermore the modifications of the momentum spectra due to interactions in the hadronic phase are well described by the model calculations. A large fraction of low momentum resonances become non-reconstructable due to the final state interactions of their decay daughters. This leads to a higher mean transverse momentum measurement for short lived resonances. 
Resonances show different centrality dependences in their yields depending on their lifetime and the interaction cross-sections of their decay products. The more central collisions show a larger contribution of hadronic phase interactions. Therefore it is important to understand this contribution since resonances make large contributions to the yields of final-state particles and to the correlations between them.

%Resonances show different centrality dependences in their yields depending on their lifetime and the interaction cross-sections of their decay products.  (Note: can we actually give the cross-sections in the paper: some measured cross-sections and/or the values used in UrQMD?)

\begin{acknowledgments}

The authors would like to thank J. Aichelin for his fruitful discussions and valuable contributions. This work was supported by U.S. Department of Energy Office of Science under contract number DE-SC0003892. This work was supported by GSI and the Hessian initiative for excellence (LOEWE) through the Helmholtz International Center for FAIR (HIC for FAIR). The authors acknowledge the Texas Advanced Computing Center (TACC) at the University of Texas at Austin for providing computing resources that have contributed to the research results reported within this paper. URL: http://www.tacc.utexas.edu. Additional computational resources were provided by the LOEWE Frankfurt Center for Scientific Computing (LOEWE-CSC).

\end{acknowledgments}


\begin{thebibliography}{99}

%Lattice


\bibitem{lattice_wu}
Y. Aoki \textit{et al.}, Phys. Lett. {\bf B643}, 46 (2006) 

\bibitem{lattice_bnl}
%Lattice  T= 155 +-9
A. Bazavov \textit{et al.}, Phys. Rev. {\bf D85} 054503 (2012) 

\bibitem{Markert_Vitev}
Markert C, Bellwied R and Vitev I, ``Formation and decay of hadronic resonances in the QGP,'' Phys. Lett. B {\bf 669}, 92--7 (2008).


%%%  EPOS3

\bibitem{epos1}
H.J. Drescher, M. Hladik, S. Ostapchenko, T. Pierog, K. Werner.
hep-ph/0007198. 
Phys.Rept. 350 (2001) 93-289.

\bibitem{epos2}
K. Werner, Iu. Karpenko, T. Pierog, M. Bleicher, K. Mikhailov.
arXiv:1004.0805 [nucl-th].
Phys.Rev. C82 (2010) 044904.

\bibitem{epos3}
K. Werner, B. Guiot, Iu. Karpenko, T. Pierog. 
arXiv:1312.1233 [nucl-th].
Phys.Rev. C89 (2014) 6, 064903.

\bibitem{epos4}
Klaus Werner. 
arXiv:0704.1270 [nucl-th].
Published in Phys.Rev.Lett. 98 (2007) 152301.



%%% URQMD
\bibitem{Bass:1998ca}
  S.~A.~Bass {\it et al.},
  %``Microscopic models for ultrarelativistic heavy ion collisions,''
  Prog.\ Part.\ Nucl.\ Phys.\  {\bf 41}, 255 (1998)
  [Prog.\ Part.\ Nucl.\ Phys.\  {\bf 41}, 225 (1998)]
  [arXiv:nucl-th/9803035].
  %%CITATION = PPNPD,41,225;%%

\bibitem{Bleicher:1999xi}
  M.~Bleicher {\it et al.},
  %``Relativistic hadron hadron collisions in the ultrarelativistic quantum
  %molecular dynamics model,''
  J.\ Phys.\  {\bf G25}, 1859 (1999)
  [arXiv:hep-ph/9909407].
  %%CITATION = JPAGA,G25,1859;%%



\bibitem{alice_reso_pbpb} B. Bezverkhny Abelev \textit{et al.} (ALICE collaboration), Phys. Rev. C{\bf 91} 2, 024609 (2015)


%High-pT resonances as a possibility to explore hot and dense nuclear matter
\bibitem{urqmd_highpt} 
S. Vogel, J. Aichelin, and M. Bleicher, Phys. Rev. C {\bf 82}, 014907 (2010)



%\bibitem{EPOS3}Particle production in proton-proton and deuteron-gold collisions at RHIC
%Klaus Werner (SUBATECH, Nantes), Fuming Liu (Hua-Zhong Normal U.), Tanguy Pierog (Karlsruhe, Forschungszentrum). Nov 2004. 4 pp. Published in J.Phys. G31 (2005) S985-S988 DOI: 10.1088/0954-3899/31/6/043 e-Print: hep-ph/0411329 | PDF 

% UrQMD

\bibitem{eposcore} K. Werner, Phys. Rev. Lett. 98, 152301 (2007)


\bibitem{cooperfrye}
F.~Cooper and G.~Frye,
Phys.\ Rev.\ D {\bf 10}, 186 (1974).

 \bibitem{Graef:2014mra} 
  G.~Graef, J.~Steinheimer, F.~Li and M.~Bleicher,
  %``Deep sub-threshold $\Xi$ and $\Lambda$ production in nuclear collisions with the UrQMD transport model,''
  Phys.\ Rev.\ C {\bf 90}, 064909 (2014)

\bibitem{urqmd_high_pt} S. Vogel, J. Aichelin and M. Bleicher, J. Phys. G: Nucl. Part. Phys. {\bf 37}, 094046 (2010)


\bibitem{urqmd_sis} S. Vogel, H. Petersen, K. Schmidt , E. Santini, C. Sturm, J. Aichelin and M. Bleicher,  Phys. Rev. C {\bf 78}, 044909 (2008)

\bibitem{urqmd_2002} M. Bleicher and J. Aichelin, Phys. Lett. B {\bf 530}, 81 (2002)

\bibitem{urqmd_2003}  M. Bleicher Nucl. Phys. A {\bf 715}, 85 (2003)

\bibitem{urqmd_2004}  M. Bleicher and H. Stoecker, J. Phys. G: Nucl. Part. Phys. {\bf 30} S111 (2004)

 \bibitem{urqmd_2005}   S. Vogel and M. Bleicher, arXiv:nucl-th/0505027 (2005)

\bibitem{urqmd_2006}    S. Vogel and M. Bleicher, Phys. Rev. C {\bf 74} 014902 (2006)

\bibitem{urqmd_2008}   S. Vogel and M. Bleicher, Phys. Rev. C {\bf 78} 064910 (2008)
 
\bibitem{Steinheimer:2015msa} 
  J.~Steinheimer and M.~Bleicher,
  %``Hadron resonance production and final state hadronic interactions with UrQMD at LHC,''
  EPJ Web Conf.\  {\bf 97}, 00026 (2015).
  
 \bibitem{Steinheimer2012}
  J.~Steinheimer, J.~Aichelin and M.~Bleicher,
 EPJ Web Conf.\ {\bf 36}, 00002 (2012).
 
 \bibitem{PDG} K. A. Olive \textit{et al.} (Particle Data Group), Chin. Phys. C \textbf{38}, 090001 (2014)

 
%Performance of the ALICE VZERO system
%CERN-PH-EP-2013-082
%Cite as: 	arXiv:1306.3130 [nucl-ex]
%\bibitem{centrality1} CERN-PH-EP-2013-082, arXiv:1306.3130 [nucl-ex]

\bibitem{centrality1} E. Abbas \textit{et al.} (ALICE Collaboration), J. Inst. \textbf{8}, P10016 (2013)


\bibitem{centrality} B. Abelev \textit{et al.} (ALICE collaboration), Phys. Rev. C{\bf 88}, 044909 (2013)

%K?(892)0 and ?(1020) production in Pb-Pb collisions at \sqrt{s}{NN}}=2.76 TeV 
%Published in Phys.Rev. C91 (2015) 2, 024609


%% 
\bibitem{ALICE_multiplicity} K. Aamodt \textit{et al.} (ALICE Collaboration), Phys. Rev. Lett. \textbf{106}, 032301 (2011)


\bibitem{blastwave_pbpb} ALICE Collaboration (Betty Abelev (LLNL, Livermore) et al.), Phys.Rev. C88 044910 (2013) 


\bibitem{thermal_raf}  M. Petran \textit{et al.}, Phys. Rev. C{\bf 88}, 034907 (2013)



\bibitem{Lin} Z.W. Lin \textit{et al.}, Phys. Rev. Lett. {\bf 89},152301 (2002)

\bibitem{Graef2012} G.~Graef \textit{et al.}, Phys. Rev. C {\bf 85}, 044901 (2012)




\bibitem{ALICE_HBT_2011} K. Aamodt \textit{et al.} (ALICE Collaboration), Phys. Lett. B \textbf{696}, 328 (2011)

\bibitem{Lisa_FemtoscopyReview} M. A. Lisa, S. Pratt, R. Soltz, and U. Wiedemann, Annu. Rev. Nucl. Part. Sci. \textbf{55}, 357 (2005)


\bibitem{markert1} B.I. Abelev et al., (STAR Collaboration), Phys. Rev. Lett. 97, 132301 (2006)

%ÒMulti-strange baryon production at mid-rapidity in Pb-Pb collisions at $\sqrt{s_{NN}}=2.76$~TeV,Ó 
\bibitem{alice_omega} 
B. Abelev et al. (ALICE Collaboration), Phys. Lett. B \textbf{728}, 216-227 (2014)

%Which has the following erratum: ÒCorrigendum to `Multi-strange baryon production at mid-rapidity in Pb-Pb collisions at $\sqrt{s_{NN}}=2.76$~TeV' 
B. Abelev et al. (ALICE Collaboration), [Phys. Lett. B 728 (2014) 216-227],Ó Phys. Lett. B \textbf{734}, 409-410 (2014).


\end{thebibliography}
\end{document}